# Emergent Complexity in a Light-Driven Self-Oscillatory Crystal: A Molecular Perspective on Autonomous Behavior and Stimulus-Modulated Motion

Yoshiyuki Kageyama,*[†] Yasuaki Kobayashi,[‡] Makiko Matsuura,[†] Toshiaki Shimizu,[†] Norio Tanada,[†] and Daisuke Yazaki[†]

**ABSTRACT:** Living organisms are molecular systems with self-sustained dynamics via energy conversion through molecular cooperation, resulting in highly complex macroscopic behaviors. Construction of such autonomous macroscopic dynamics at a molecular system level remains one of the central challenges in the field of chemistry. Looking further ahead, constructing motile systems that can receive external information and adapt their autonomous behavior represents the next frontier towards newly functional molecular devices such as microrobots. In this study, we focused on a light-driven self-oscillatory crystal that exhibits continuous flipping motion under constant light irradiation. We experimentally evaluated the oscillation frequency of the crystal under polarized light and confirmed the validity of our previously proposed mechanism, and clarified the requirements for self-oscillation. Based on this mechanism, we constructed a mathematical model that demonstrates the motion of the crystal itself. The model revealed that diverse oscillatory behaviors at the macroscopic level can emerge due to small differences in conditions even when the underlying molecular-level processes are identical. Furthermore, we found that the oscillatory behavior depended on the state generated by the previously applied light. This suggests that a memory effect contributes to the complex motion of the crystal.

## Introduction

The bottom-up construction of molecular machines to create a life-like autonomous architecture represents a longstanding challenge in the fields of molecular science and nanotechnology. [1-4] A fundamental characteristic of such architectures is their self-repetitive mechanical work at the macroscopic level, [5] as well as in their potential for self-replication and intelligent behavior. [6] An inanimate chemical system will generally achieve a balanced state of chemical reactions, known as equilibrium or a stationary state, in which the system maintains a static composition determined by kinetic coefficients. [7, 8] For example, under stimulus application, stimulus-responsive materials are transiently deformed to the balanced state. Therefore, for repetitive demonstration of a deformation, researchers usually repeat stimulus applications with alternating reaction velocities. [9-13] In contrast, some researchers have successfully exhibited continuous dynamics of stimulus-responsive materials without modulation of the stimulus. One of these methods is the application of a self-shadowing mechanism, in which the receiving rate of the stimulus is self-modulated by the deformation of the subjected material. By employing this mechanism, White et al. developed high-frequency oscillation of an azobenzene elastomer immobilized on an object, [14] and Broer et al. demonstrated directional crawling of a frame-fixed azobenzene polymer. [15] Numerous studies using this highly versatile mechanism have been reported to date. [16-33] Similarly, continuous dynamics through the switching of heat flow caused by object deformation have been reported. [34-36] The key for the emergence of such continuous dynamics is the position of the energy source relative to the deformation form; therefore, it remains uncertain whether continuous motion can be achieved when energy is incident from all directions. [37] This raises an important point: although energy is a scalar quantity, the expression of autonomy often depends on directional interactions. Notably, there are systems in which autonomy is achieved through energy conversion independent of the relative position of the energy sources. Chemically oscillating hydrogels illustrate such behavior. [38, 39] One important advancement in this area is the construction of a self-oscillating gel system with a constant period independent of temperature that combines the functions of a chemically oscillating gel and a thermo-responsive gel. [40] In parallel, the bottom-up construction of molecular assemblies with continuous dynamics has been realized via autocatalytic processes with time-delayed feedback at both the flask level[41-43] and the molecular assembly level, [44, 45] despite the absence of mechanical work.

Previously, we reported the light-driven self-sustained oscillation of a cocrystal composed of the azobenzene derivative 6-[4-(phenylazo)phenoxy]hexanoic acid (**1**) and oleic acid in a ~3:2 molar ratio, demonstrating an example of the nonself-shadowing type of light-driven self-oscillation. [46-48] In our first paper, we revealed that the self-sustaining oscillation, the frequency of which is proportional to the incident light intensity, is realized by the fact that the photoisomerization-triggered crystalline phase transition causes switching of the kinetic coefficients of the photoisomerization in the crystal (Fig. 1). [46] Similar self-oscillations, but with more nondeterministic behavior, also have been observed in the single-component assembly of **1**[47] and co-assemblies of some azobenzene derivatives with fatty acids,[49] thus indicating that the phase transition involving light-driven self-oscillation is not a rare phenomenon. An advantage of this mechanism is that conditions related to relative directionality, such as the position of the energy source, are not a critical requirement,[37] unlike the other reported continuously moving photomechanical materials working under directional light.[14-28, 35] Therefore, regardless of whether it adopted a "standing" or "lying" orientation with respect to the incident light, our crystal continued to flip (Movie S1 in the Electronic Supporting Information (ESI)). Furthermore, the underwater self-propulsion of the crystal arose from mechanical forces acting on the surrounding medium and overcame Purcell's theorem, which states that small flappers cannot swim in water. This propulsion was determined not by the incident light direction but by its anisotropic shape during deformation.[48] Conversely, when a directional energy source was used, the original self-sustaining oscillation of the object was disturbed because the reaction probability was affected by the directionality. Indeed, we have previously reported that the shapes of the repetitive motion vary in response to changes in the polarization angle.[47] In that paper, we also revealed the single-crystal structure of a polymorph of **1**, whose powder X-ray diffraction pattern was similar to that of the cocrystal in the as-synthesized phase. According to the crystal structure with six crystallographically independent azobenzene molecules, we concluded that the cocrystal consisted of four crystallographically independent azobenzene molecules and two crystallographically independent oleic acid molecules (Fig. 2), whereby one of the azobenzene molecule acts as the oscillation generator while the other azobenzene molecules regulate the flipping properties.

In the present study, the frequency of self-oscillation of the same cocrystal under polarized light was examined, and the origin of the polarized-light-triggered variation in the frequency was determined with the support of a model derived from the reaction kinetics. Thus, we established the validity of the proposed mechanism that the isomerization reaction is reversed by the phase transition, as suggested in our previous paper.[46] In addition, by extending the model to crystalline structures, we demonstrated one reason why real crystals exhibit diversity despite the same mechanisms operating at the molecular level. Furthermore, we discovered a novel type of hysteresis in autonomous dynamics due to the "remembering" information from the previous external operation. These characteristics of multimolecular cooperative systems were realized through the combination of autonomous systems and nonautonomous molecular behaviors. We hope that this paper will serve as a starting point for considering what it truly means for us, as chemists, to pursue the construction of autonomous molecular systems.

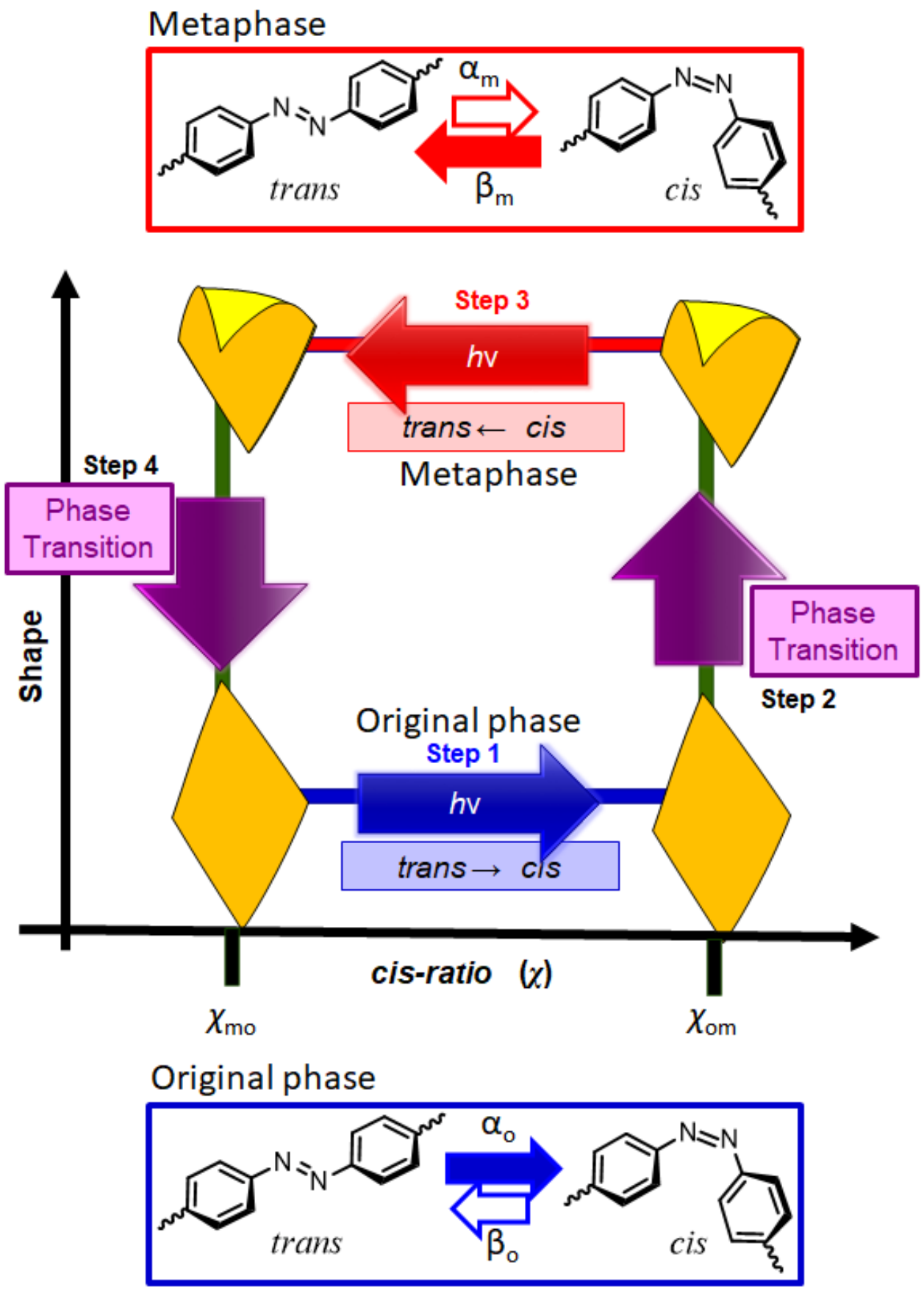

**Figure 1.** Schematic illustration of the proposed mechanism of limit-cycle self-oscillation of the cocrystal composed of **1** and oleic acid under continuous blue-light irradiation.[46]

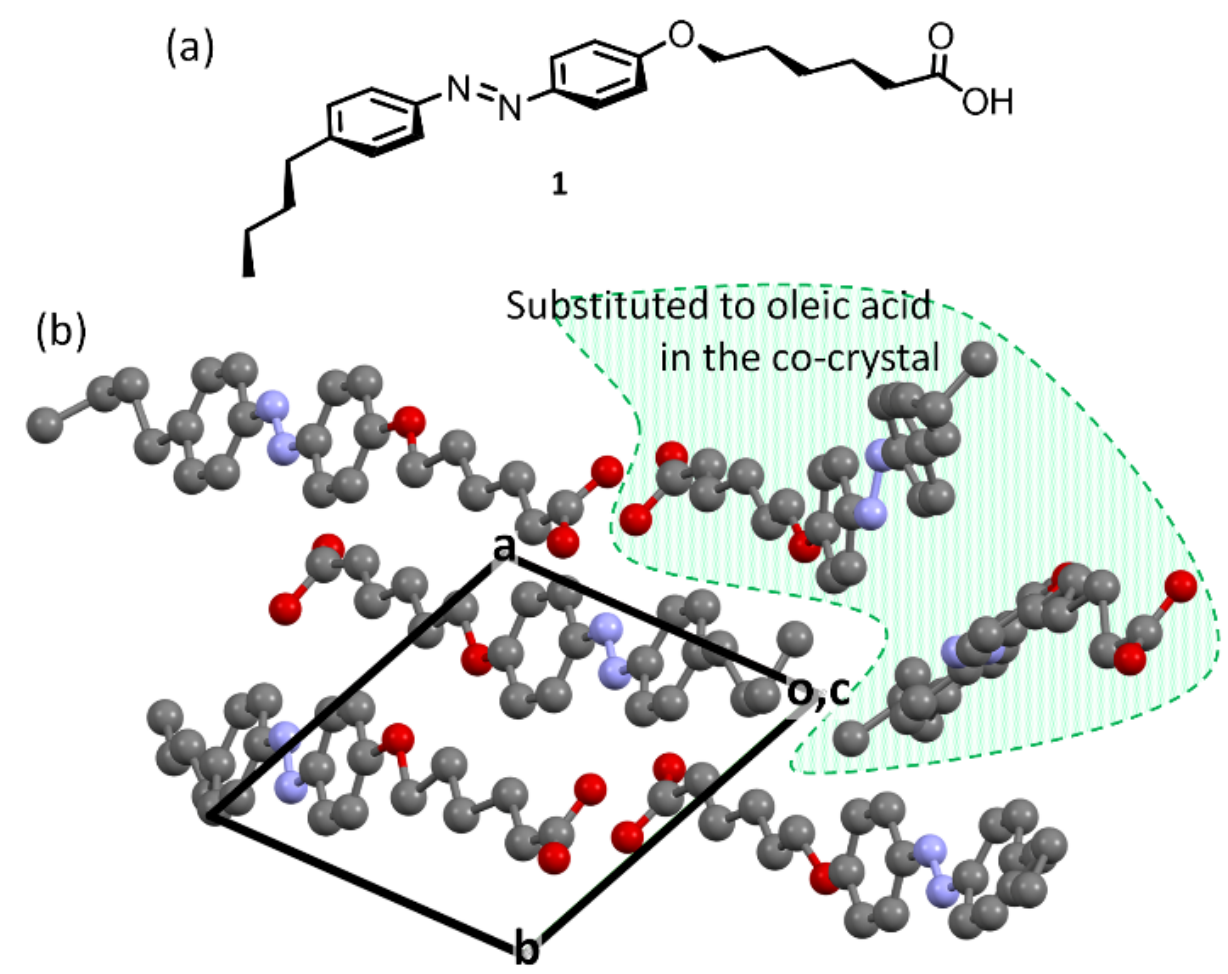

**Figure 2.** Schematic illustration of (a) the chemical formula of **1** and (b) the unit cell of single crystals of **1**, which crystalizes in the space group $P_1$ (the crystallographic data is available from the Cambridge Crystallographic Data Centre (CCDC183362)); the hydrogen atoms in **1** have been omitted for clarity. The structure of the cocrystals between **1** and oleic acid was tentatively assigned to be the same as that illustrated above, except for the two encircled molecules of **1**, which lack effective dipole-dipole interactions around the ether oxygen and are thus presumed to be replaced by oleic acid.[47]

## Methods

### *Preparation of the cocrystal*

Cocrystals of **1** and oleic acid were prepared according to a literature procedure, [47] and the compound identification data matched those reported previously. [50] In brief, a mixed dispersion of 1 mg of the *trans* isomer of 6-[4-(phenylazo)phenoxy]hexanoic acid (**1**) and 1.3 mg of sodium oleate in 1 mL of phosphate-buffered solution (pH = 7.5; 75 mM) was ultrasonicated at 55 °C for 0.5 h and incubated at 25 °C for more than one day to obtain a dispersion of cocrystals. The dispersion was placed on a glass slide and sealed for observation.

### *Observation method*

For microscopic observation, an inverted fluorescence microscope (TE2000 system, Nikon, Japan) equipped with an epifluorescence module (C-HGFI, Nikon, Japan) for light illumination from the bottom side of the sample was used. This microscope consisted of a precentered fiber illuminator with a mercury lamp, a turret for filter units for light excitation, a polarizer, and a Nikon Plan Fluor ELWD 20× (NA0.45) objective lens. In the turret, Nikon filter cube units, UV-1A ($\lambda_{ex}$ = 365 ± 5 nm), V-1A ($\lambda_{ex}$ = 400 ± 20 nm), BV-1A ($\lambda_{ex}$ = 435 ± 10 nm), B-2A ($\lambda_{ex}$ = 470 ± 20 nm), B-3A ($\lambda_{ex}$ = 455 ± 35 nm), and G-2A ($\lambda_{ex}$ = 535 ± 25 nm), were used. The excitation photon number was $6 \times 10^{15}$/s for polarized light and $2 \times 10^{16}$/s for unpolarized light, as measured using a power meter (LP1 Mobiken, Sanwa, Japan), and the light intensity was controlled using neutral density filters and a neutral density-control dial on the epifluorescence module. As the second light source, a custom-made light-excitation unit was added to the upper side of the microscope, as shown in Fig. S1 in ESI. An IR cut-off filter, a 435-nm bandpass filter (ASAHI Spectra, Japan), and a 455-nm dichroic mirror (provided by Nikon as a part of the BV-1A filter unit) were equipped to the unit. The second light source was used in experiments in which the light irradiation conditions were switched during the course of the observation. The observations were conducted using either the polarization microscopy method or the differential interference contrast microscopy method. A charge-coupled device camera (STC-TC152USB, Sentech, Japan) was used to record movies at 19.26 fps, and Image-Pro Premier software (Media Cybernetics, USA) was used for movie analysis. When more than one oscillation mode was present, the frequency was measured for the long-period oscillation mode. The frame rate of the movies in ESI were processed to be 23.98 fps by using Adobe Premier Pro software (USA).

### *Model for the self-oscillation frequency based on reaction rate equations ignoring the time required for phase transition*

Using the expression for reversible photoisomerization between the *trans* isomer and the *cis* isomer of azobenzene (eq (1)), the coarse-grained reaction rate eq (2) and eq (3), wherein thermal processes are ignored, can be established.

$$\textit{trans isomer} \underset{\beta}{\overset{\alpha}{\rightleftarrows}} \textit{cis isomer} \tag{1}$$

$$\frac{\mathrm{d}}{\mathrm{dt}}\chi = \alpha - (\alpha + \beta)\chi \tag{2}$$

$$\Delta\chi = \int_{t_0}^{t_1} \frac{d}{dt}\chi \, dt = \int_{t_0}^{t_1} (\alpha - (\alpha + \beta)\chi) \, dt \tag{3}$$

In these expressions, α and β are the kinetic coefficients of the *trans*-to-*cis* and *cis*-to-*trans* photoisomerization, respectively, χ is the fraction of the *cis* isomer, and Δχ is the difference in χ between time $t_0$ and time $t_1$. Each kinetic coefficient is multiplied by the quantum yield of isomerization (Φ), the excitation probability, and the light intensity ($I$). Generally, the excitation probability is proportional to $|\boldsymbol{M}|^2 \cos^2\Delta\theta$.[51] Therefore, α and β can be expressed as shown in eq (4) and eq (5):

$$\alpha = C \times \Phi_t \, |\boldsymbol{M_t}|^2 \cos^2(\theta_t - \theta_\mathrm{P}) \; I = A\cos^2(\theta_t - \theta_\mathrm{P}) \tag{4}$$

$$\beta = D \times \Phi_c \, |\boldsymbol{M}_\mathbf{c}|^2 \cos^2(\theta_c - \theta_\mathrm{P}) \; I = B\cos^2(\theta_c - \theta_\mathrm{P}) \tag{5}$$

where $\theta_\mathrm{p}$ is the polarization angle of the light, which is irradiated from the vertical direction, θ is the azimuth angle of the transition moment, $I$ is the incident light intensity, and $A$–$D$ are constants that depend on the elevation angle of $\boldsymbol{M}$. The subscripts $t$ and $c$ indicate the *trans* isomer and *cis* isomer, respectively. Since this paper does not discuss light intensity changes, $I$ is always set to 1.

In the self-oscillating crystal with the mechanism shown in Fig. 1, the kinetic parameters differ between the original phase and the metaphase. Herein, we consider the behavior of the generator molecule. The phase transitions are triggered when χ, the *cis*-isomer fraction, reaches the respective threshold ($\chi_{om}$ and $\chi_{mo}$ for the original-phase-to-metaphase transition and the metaphase-to-original-phase transition, respectively). When $DT_\mathrm{o}$ and $DT_\mathrm{m}$ are defined as the duration times of the original phase and the metaphase, respectively, $\Delta\chi_{tt}$, which is the difference in χ between the two thresholds, can be expressed as shown in eq (6) and eq (7), which are derived from eq (3):

$$\Delta\chi_{tt} = \int_0^{DT_\mathrm{o}} \frac{\mathrm{d}}{\mathrm{d}t}\chi \, \mathrm{d}t = \int_0^{DT_\mathrm{o}} (\alpha_\mathrm{o} - (\alpha_\mathrm{o} + \beta_\mathrm{o})\chi) \, \mathrm{d}t \tag{6}$$

$$-\Delta\chi_{tt} = \int_0^{DT_\mathrm{m}} \frac{\mathrm{d}}{\mathrm{d}t}\chi \, \mathrm{d}t = \int_0^{DT_\mathrm{m}} (\alpha_\mathrm{m} - (\alpha_\mathrm{m} + \beta_\mathrm{m})\chi) \, \mathrm{d}t \tag{7}$$

where the subscripts o and m indicate the original phase and the metaphase, respectively. The periodic time is the sum of $DT_\mathrm{o}$ and $DT_\mathrm{m}$ if the phase transitions occur quickly enough; thus, the frequency ($F$) can be expressed as eq (8):

$$F = \left(\frac{1}{DT_o + DT_m}\right) \tag{8}$$

### *Mathematical four-dimensional model representing the light-driven self-oscillation with introduction of the first-order phase-transition model*

To reproduce the crystal structure, hexagonal structures with triangular lattices were employed (Fig. 3). [52, 53] The numbers of nodes forming the three sides of the hexagonal structure were (6, 6, 29), (8, 8, 21), (9, 9, 17), or (10, 10, 15), with the total numbers of the nodes being 333, 349, 336, or 347, respectively. The fraction of metaphase ($\phi_i$), *cis*-isomeric ratio ($\chi_i$), and coordinates ($r_i$) were defined in each node, with the time evolutions calculated using the equations shown below. In the numerical simulations, each triangle in the crystal was colored according to the values of these variables averaged over the three nodes of the triangle.

Reaction velocities

In the present model, we introduced the following exponential relationships between the kinetic coefficients and the metaphase fraction ($\phi_i$), which is defined between 0 and 1:

$$\alpha(\phi_i) = 7.5 \times 10^{-5} \times \exp(-5\phi_i) \times CS \times I \qquad (9)$$

$$\beta(\phi_i) = 7.5 \times 10^{-6} \times \exp(5\phi_i - 0.5) \times CS \times I \qquad (10)$$

In this study, it was assumed that collimated light is emitted from the upper side of the object (in the Z-axis direction). In this case, the light-receiving area was the projected area of the triangle from the Z-axis direction to the XY plane. $CS$ is the factor that adjusts the change in the light-receiving area of each triangle depending on the direction of light irradiation. In this study, we assumed that parallel light is irradiated from the top side of the objects. It is noteworthy that stable oscillations will occur even if light irradiation is assumed to be isotropic, i.e., $CS = 1$. (Although this case is not discussed in this paper, the results under this condition are shown in Movie S8.) Using these kinetic coefficients, eq (2) for the *cis*-isomeric ratio $\chi_i$ is modified as

$$\frac{d\chi_i}{dt} = \alpha(\phi_i) - \left(\alpha(\phi_i) + \beta(\phi_i)\right)\chi_i. \qquad (11)$$

Landau theory for the first-order phase transition

Referring to the mean-field model of Landau theory, we employed eq (12) and eq (13) for the pseudo energy function ($f_i$) to indicate the stability of the phase.

$$f_i(\phi_i, \chi_i) = \frac{\Lambda(\chi_i)}{2}\phi_i^{\,2} - \frac{B}{4}\phi_i^{\,4} + \frac{C}{6}\phi_i^{\,6} \qquad (12)$$

$$\Lambda(\chi_i) = L \times (\chi_{\mathrm{PT}} - \chi_i) \qquad (13)$$

In the present study, we employed 1.5, 0.3, 1.6, and 2.0 for the quantities of $L$, $\chi_{\mathrm{PT}}$, $B$, and $C$, respectively. We also considered the diffusive interaction between the adjacent nodes, with the interaction strength $D_\phi = 0.01$ and the slowness of the dynamics represented by the parameter $\varepsilon_\phi = 0.01$. $\phi_i$ is given by eq (14):

$$\frac{1}{\varepsilon_\phi}\frac{d\phi_i}{dt} = D_\phi \sum_j (\phi_j - \phi_i) - \frac{\partial f_i}{\partial \phi_i}. \qquad (14)$$

Mechanical interaction between nodes

For each node, we assumed that the spontaneous curvature at each edge of the lattice depends on the metaphase fraction $\phi_i$. Assuming the stretching energy between two nodes and the bending energy between two triangles, the change in coordinates of node $r_i$ is governed by eq (15):

$$\frac{d\boldsymbol{r}_i}{dt} = -\frac{\partial}{\partial \boldsymbol{r}_i}(V_{str} + V_{bend}) \qquad (15)$$

$$V_{str} = \frac{K_{spring}}{2}\sum_{(i,j)\in edges}\left(\left|\boldsymbol{r}_i - \boldsymbol{r}_j\right| - l_0\right)^2 \qquad (16)$$

$$V_{bend} = K_{bend}\sum_{k\in edges}(1 - \cos(\theta_k - \psi_k)) \qquad (17)$$

where the summation in $V_{str}$ is taken over the pairs of nodes $(i, j)$ forming edges. The summation in $V_{bend}$ is taken over all edges, where $\theta_k$ is the bending angle between the two adjacent triangles sharing edge $k$, and $\psi_k = 0.1(\phi_i + \phi_j)/2$ is the stable angle determined by the spontaneous curvature, which is assumed to be proportional to the average $\phi$ values at the two nodes $i$ and $j$ forming edge $k$. The lattice constant $l_0$ is set to 1.0. The spring constants for these potentials ($K_{\mathrm{spring}}$ and $K_{\mathrm{bend}}$, the values of which are set to 20 and 8, respectively) were also defined. In addition, to reproduce the fact that the bending was parallel to a side of the crystal, we introduced an asymmetric factor into the bending rigidity in such a way that all edges parallel to a given axis (indicated by the arrow in Fig. 3) were assigned the half value of $K_{\mathrm{bend}}$.

In mathematical models, parameters are inherently relative. In this study, we selected parameter sets that enhance the clarity of behavioral trends in visualized simulations and minimize discretization errors. It is noteworthy that the use of different parameter values from those in the previously introduced model does not affect the essence of the discussion. This study aimed to model and understand the experimental results rather than to optimize the parameters for the exact reproduction of those results. The number of calculated steps is $4 \times 10^5$, and the value of time is $4 \times 10^3$, in general, with time step $\Delta t = 0.01$.

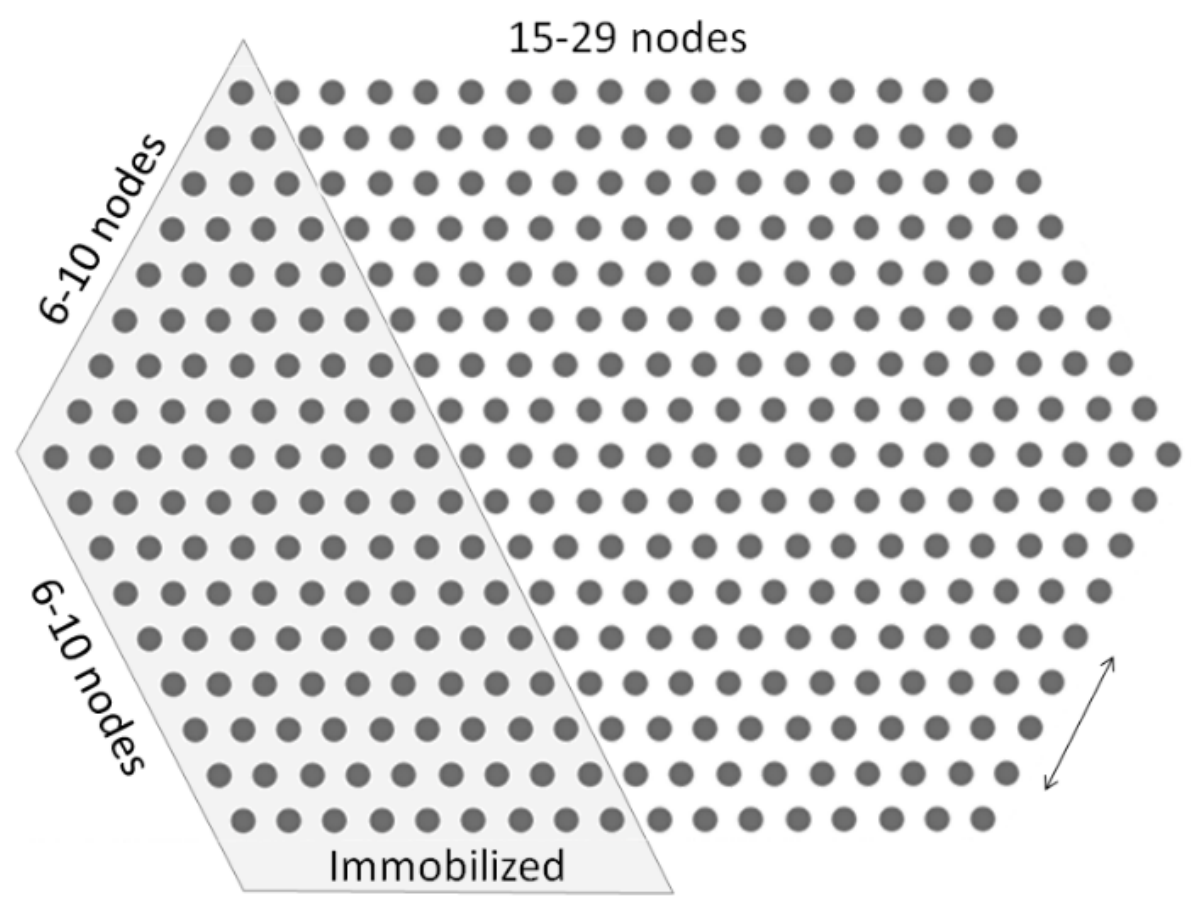

**Figure 3.** The crystal model used in the numerical calculations. The isomer fraction, phase-state fraction, and spontaneous curvature at each node (shown as ●) as well as the elastic interactions between the nodes were calculated. The value of $K_{\mathrm{bend}}$ was halved for edges parallel to the arrow.

## Results and Discussion

*Light-polarity dependence of the self-oscillation*

A crystal composed of the *trans* isomer of **1** and oleic acid was placed on a glass slide as an aqueous dispersion. The excitation light ($\lambda_{\mathrm{ex}}$ = 435 nm) was applied vertically (i.e., perpendicular to the face of the glass slide), and its polarization azimuth was defined with respect to the micrographic view. The azimuth angles presented in this paper as experimental data were modified by the corresponding angles in case the crystals rotated during the experiment. The crystals oscillated in the vertical direction; the oscillation was detected as the repeated observation of an in-focus flat object and an out-of-focus bent object, or as the brightness change of the object. In the experiment, we first exposed the crystal to unpolarized light ($\lambda_{\mathrm{ex}}$ = 435 nm) for 5 s to generate the self-oscillation, and then we measured the oscillation frequency under polarized light ($\lambda_{\mathrm{ex}}$ = 435 nm). The frequency of the resulting self-oscillation under polarized blue-light irradiation was visualized as a polar plot, with the azimuth angle of the polarized light indicated.

Examples of the results are shown in Fig. 4–5, and other examples are depicted in Fig. S2–S3 in ESI. Since the shape of the crystal varied greatly, the polarization-angle dependence was also different for each crystal. However, the dependencies can be classified into two groups: the two-leaf type and the four-leaf type. Generally, the excitation probability of a molecule as a function of the light polarity is proportional to $|\boldsymbol{M}|^2 \cos^2\Delta\theta$, where $\boldsymbol{M}$ is the transition moment of the molecule and $\Delta\theta$ is the differential angle between the azimuth angle of $\boldsymbol{M}$ and that of the light polarization. [51] Therefore, polar plots of light-triggered deformation of materials generally show a symmetrical petal-like shape with two lobes. However, in our experiments, the shapes of the polar plots were not proportional to $\cos^2\Delta\theta$. Regarding the two-leaf-type plots, the forbidden angular spans were wide and the line symmetry was broken compared to the line of $\cos^2\Delta\theta$. Meanwhile, regarding the four-leaf-type plots, the shapes were fully different from the shape of $\cos^2\Delta\theta$. Furthermore, even in cases in which the flipping frequency was similar for different light-polarization angles, the time profile of the flipping was different. For example, Fig. 4c presents the time profile of the flipping motion of the same crystal under 100°- and 170°-polarized light. The duration ratio of the bent shape to the flat shape (*Dr*) was 1.4 at 170° and 0.96 at 100°. Other examples are also shown in Fig. 5c and Fig. S3 in ESI.

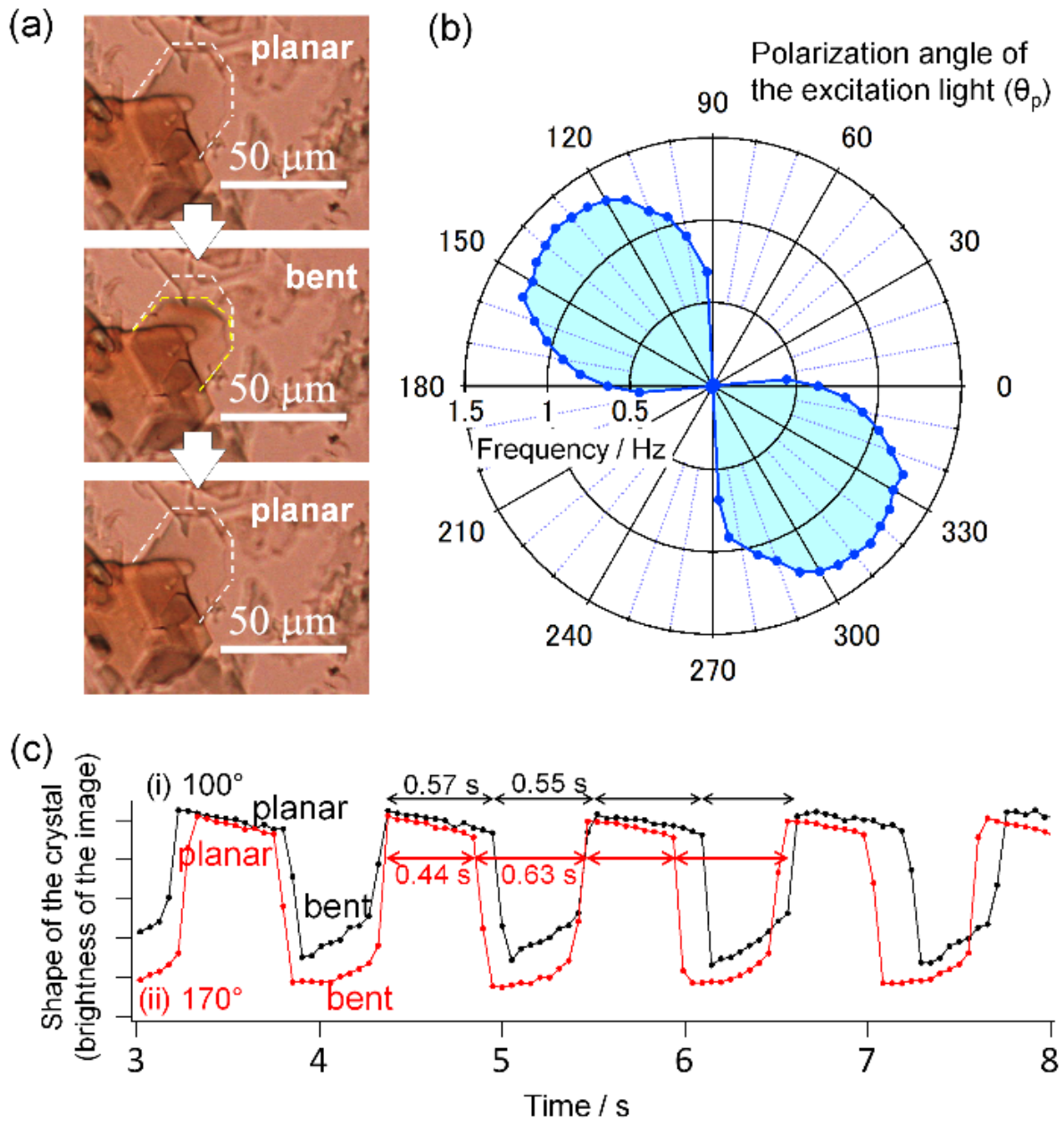

**Figure 4.** (a) Sequential microscopic images (interval: 0.42 s) of the oscillatory flipping of the crystal under blue-light ($\lambda_{ex}$ = 435 nm) excitation with a polarization angle of 140°. The dashed lines are guides for the eye. (b) Frequency of the self-oscillation of a crystal under blue-light polarization ($\lambda_{ex}$ = 435 nm), obtained after irradiation with unpolarized blue light ($\lambda_{ex}$ = 435 nm) for 5 s. The frequency resolution is approximately 0.05 Hz due to the frame speed of the video used to generate this image (for details, see Movie S2 in the Electronic Supporting Information (ESI)). (c) Time profiles of the self-oscillation under light with a polarization angle of 100° and 170°, respectively. The vertical axis is the brightness of a part of the crystal in the video used to generate this image (for details, see Movie S2 in ESI) and indicates the state of the crystal.

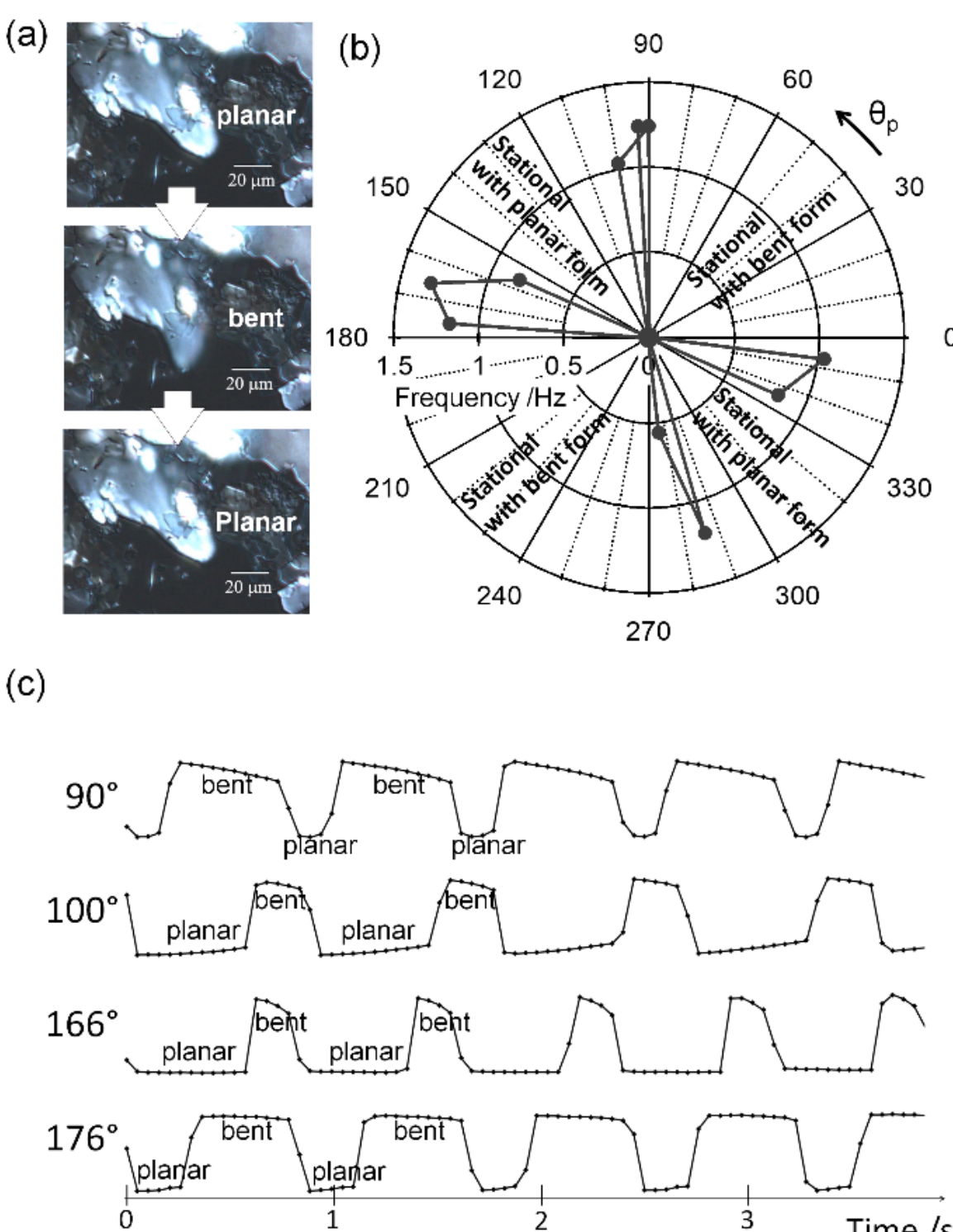

**Figure 5.** (a) Sequential microscopic images (interval: 0.36 s) of the oscillatory flipping of a tilted crystal under blue-light excitation ($\lambda_{ex}$ = 435 nm) with a polarization angle of 90°. (b) Frequency of the self-oscillation of the crystal under blue-light polarization, obtained after irradiation with unpolarized blue light ($\lambda_{ex}$ = 435 nm) for 5 s. The frequency resolution is approximately 0.05 Hz due to the frame speed of the video (for details, see Movie S5 in ESI). (c) Time profiles of the oscillations under light with a polarization angle of 90°, 100°, 166°, and 176°.

*Analysis of the model constructed from the reaction kinetic equations*

The experimental results were successfully simulated by the model based on the proposed scheme (Fig. 1), where the oscillation was realized by both the *trans*-isomer-decreasing photoisomerization of the generator azobenzene molecule at the original crystalline phase (flat shape) and the *cis*-isomer-decreasing photoisomerization of the generator azobenzene molecule at the metaphase crystal (bent shape). [46, 47] In this section, we discuss the theoretical aspects.

In general, especially in solution phases, for the reversible photoisomerization of the azobenzene derivative shown in eq (1) in the Methods section, the expected *cis*-isomeric ratio at a photostationary state (expected $\chi_{pss}$) is $\alpha/(\alpha+\beta)$. Because the kinetic coefficients, α and β, depend on the surrounding conditions of the molecule, the $\chi_{pss}$ value is shifted based on the condition, such as the dielectric parameter of the solvent. Even in the case of crystals, the kinetic coefficient varies with the surrounding conditions. In this study, we considered the photoisomerization behavior in the self-oscillatory crystal. First, we considered the *cis*-isomer-increasing process of the self-oscillatory crystal under light irradiation. If the expected $\chi_{pss}$ is less than the phase-transition threshold ($\chi_{om}$ in Fig. 1), χ never reaches $\chi_{om}$, and the subjected material never shows phase transition (Case A in Fig. 6a). In this case, the material remains at the photostationary state. On the other hand, if the expected $\chi_{pss}$ is greater than $\chi_{om}$, the material undergoes the phase transition before it reaches the photostationary state (Case B in Fig. 6a). Next, we considered the *trans*-isomer-increasing process. Because of the phase transition, the kinetic coefficients are different from those in the *cis*-isomer-increasing process. In the *trans*-isomer-increasing process, phase transition occurs only if the expected $\chi_{pss}$ is less than $\chi_{mo}$ (Fig. 6b); otherwise, the material turns into a photostationary state or an undiscussed state (Case C in Fig. 6a). In summary, as illustrated in Fig. 6c, we could predict the phase diagram showing the dynamic behavior.

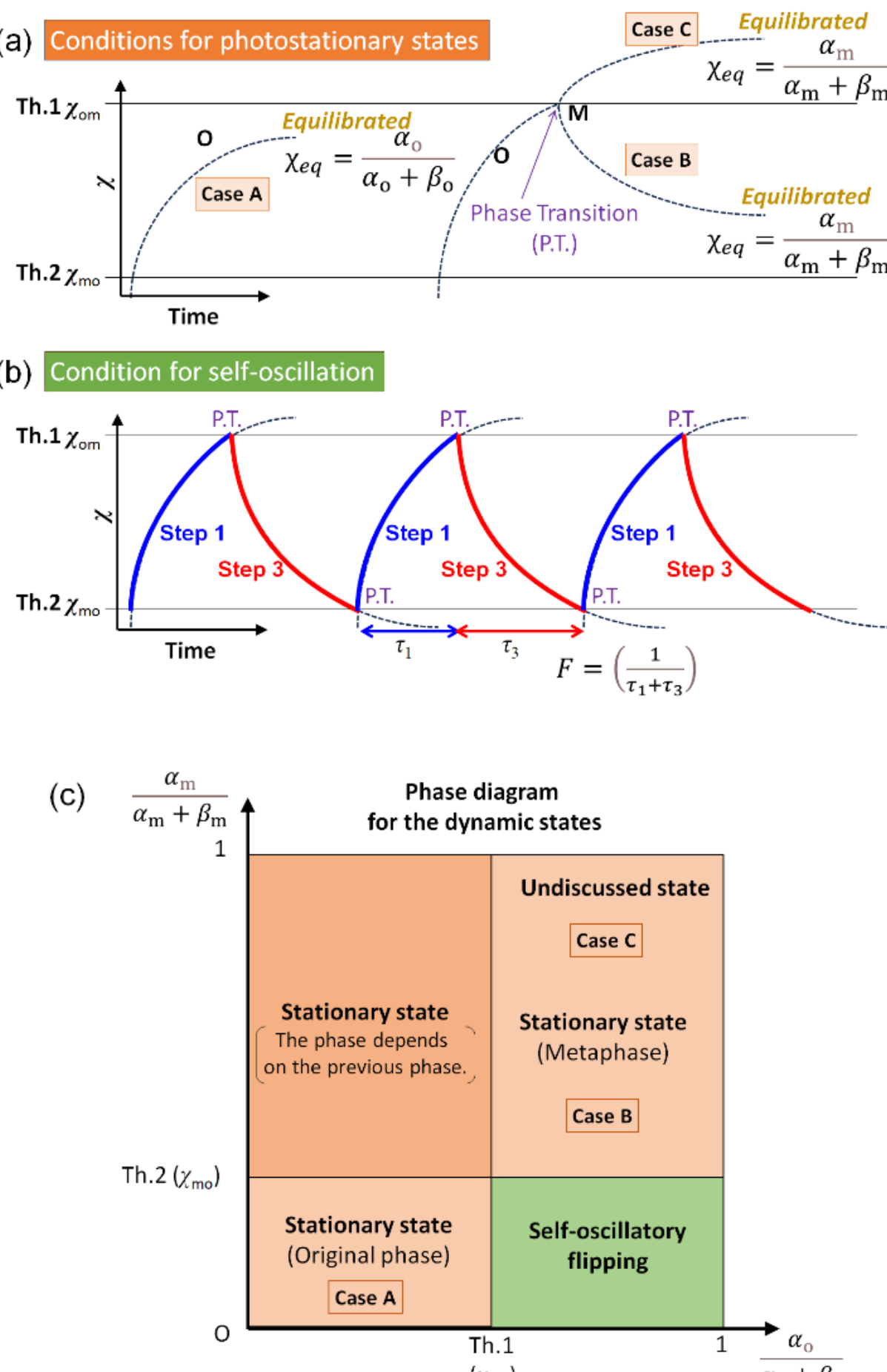

**Figure 6.** Schematic illustrations showing the conditions for light-driven self-oscillation. (a) Reaction rate conditions leading to photostationary states. (b) Reaction rate conditions leading to light-driven self-oscillation. (c) A phase diagram representing the dynamic states of the crystal based on the reaction rate conditions. Abbreviation: Th, threshold.

The values of the kinetic coefficients, α and β, also depend on the wavelength of the excitation light. Therefore, the availability of self-oscillation must depend on the excitation wavelength. Indeed, in the previously reported experiments, upon unpolarized light irradiation, we observed self-oscillatory dynamics if the light color was blue or blue-violet, while we did not observe them with irradiation of green light ($\lambda = 535 \pm 25$ nm), violet light ($\lambda = 400 \pm 20$ nm), nor ultraviolet light ($\lambda = 365 \pm 5$ nm).[46] Under green light irradiation, where $\alpha_o$ is expected to be small, we did not observe any bending of the crystal; while under violet or ultraviolet light, where $\beta_m$ is expected to be small, we observed deformation followed by melting of the crystal. Furthermore, we have reported a shift in the ratio of duration times (the ratio of $DT_o$ and $DT_m$) by changing the light wavelength.[46] In short, $DT_o/DT_m$ was larger under 470-nm light irradiation than that under 435-nm light, and the $\chi_{pss}$ values in methanolic solution were 78% and 72% for 470-nm excitation and 435-nm excitation, respectively. The experimental results are consistent with the above-mentioned theoretical discussion.

If the excitation light is polarized, α and β also depend on the relative angle of the polarization against the transition moment of the subjected molecule, as shown in eq (4) and eq (5). The expected $\chi_{pss}$ values at each phase, i.e., $\alpha_o/(\alpha_o+\beta_o)$ and $\alpha_m/(\alpha_m+\beta_m)$, should vary between 0 and 1 depending on the relative angle, which is defined between 0° and 180°. This fact means that the kinetic parameters among the change in polarization angle between 0° and 180° are displayed as a closed line from the bottom end to the top end and from the left end to the right end on the phase diagram of the dynamics, as shown in Fig. 7. If the line crosses the self-oscillatory region once, the polar plot shape is the two-leaf type; and if it crosses twice, the polar plot shape is the four-leaf type.

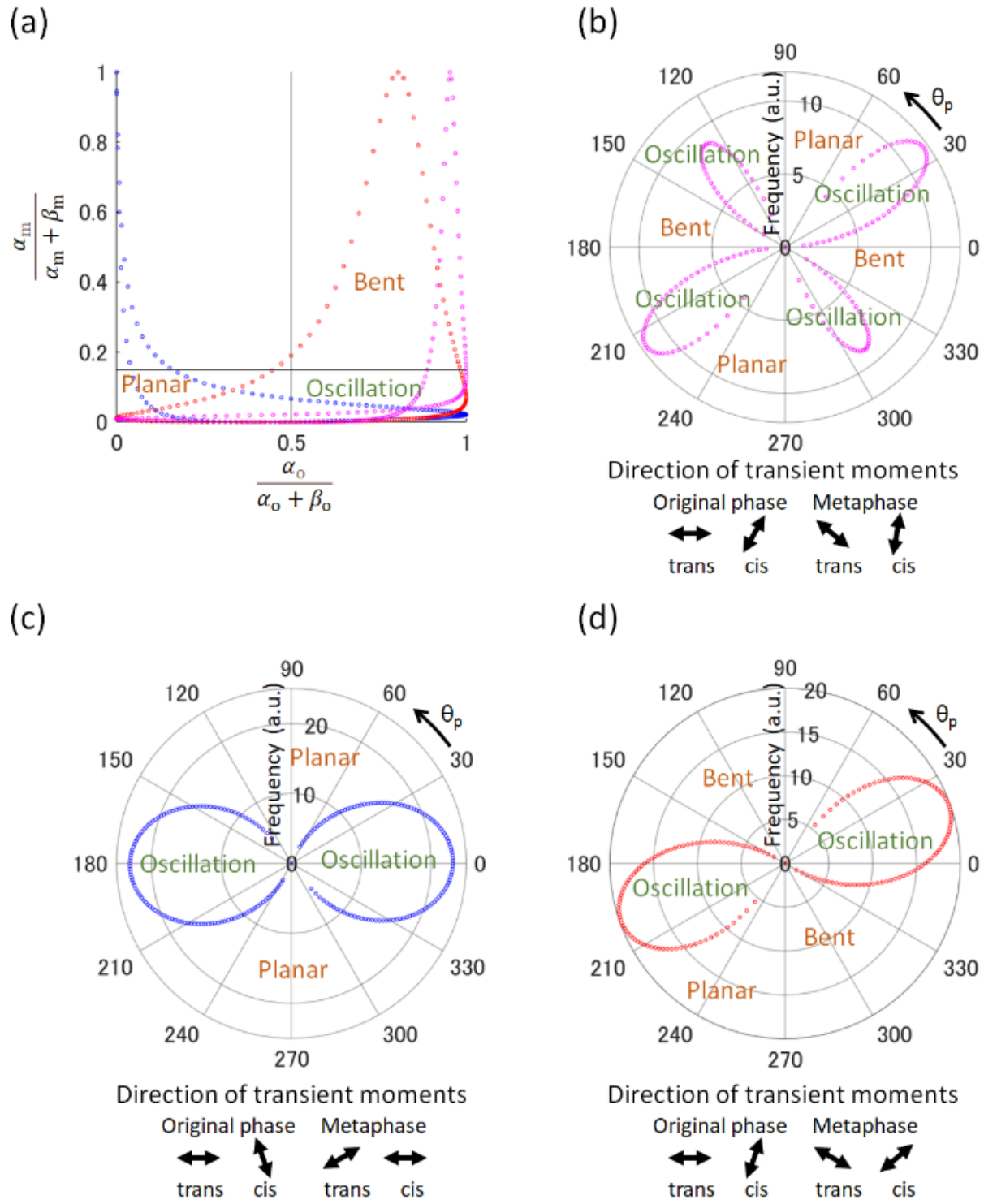

**Figure 7.** Three examples of the polarization angle dependence of self-oscillation predicted by the simple model based on the rate equations. The relative values of A, B, C, and D in eq (4) and eq (5) were 30, 12, 1, and 50, respectively. The values of $\theta_t$ and $\theta_c$ for the original and metaphases were set to [0°, 60°, -40°, 80°] (magenta), [0°, -70°, -30°, 0°] (blue), and [0°, 70°, -30°, 40°] (red), respectively. The angles are illustrated as double arrows in panels b, c, and d. Calculations were performed in 1° increments. Herein, we set $\chi_{om}$ and $\chi_{mo}$ as 0.5 and 0.15, respectively. In (a), the magenta case, which enters the self-oscillatory region twice in the phase diagram, exhibits a four-leaf polarization angle dependence as shown in (b). In contrast, the blue and red cases, which enter the region only once, are expected to show a two-leaf polarization angle dependence as shown in (c, d). From this theoretical approach, it is expected that in the case of the four-leaf type, bent and planar stationary states will be observed on opposite sides of the region of self-oscillation. On the other hand, in the case of the two-leaf type, it is predicted that the bent and planar stationary states may be adjacent to each other, or only the stationary planar state may be observed.

In terms of a qualitative discussion, the shape of the polar plot can also be understood as stated below. If the polarization angle is perpendicular to the transition moment of the *trans* isomer in the original phase, *trans*-to-*cis* photoisomerization at the original phase is prohibited. If the polarization angle is perpendicular to the transition moment of the *cis* isomer in the metaphase, *cis*-to-*trans* photoisomerization at the metaphase is prohibited. Due to the two forbidden angles, the shape becomes four leaves; and if the forbidden angles are close each other, or depending on the relationship between the threshold values and the reaction rate constants, it becomes two leaves.

The experimental results were congruent with the theoretical aspects. Not only the existence of the two-leaf- and four-leaf-type plots, but also the trends of angle-dependent shifts in the ratio of $DT_o$ and $DT_m$ were in good agreement with the theoretical discussions (Fig. S4 in ESI). In addition, the crystalline shape at the photostationary state differed before and after self-oscillation in the four-leaf case; this finding is consistent with the theoretical prediction. Similarly, in the two-leaf case, the experimental results indicated that the shape in the photostationary state switched from bent to planar or from planar to bent within the nonoscillatory angular range. These observations were also consistent with the theoretical discussion.

On the other hand, the theoretical consideration suggests that physical factors, such as shifts in the reaction rate ratios, phase-transition thresholds arising from the crystal morphology (e.g., variations in the number of layers or twin formations), and the internal crystal environment (e.g., differences in void content or temperature factors), as well as shifts in the reaction rate ratios dependent on the excitation light wavelength characteristics and angular shifts resulting from the tilt of the crystal relative to the light source may alter the self-oscillation frequency and its dependence on the polarization angle. In other words, while the molecular-level mechanism provides an explanation for the macroscopic self-oscillation, it does not allow for a unique or precise prediction of its dynamic behavior. Such nondeterministic characteristics are intrinsic to heterogeneous dynamic autonomous systems, where cooperative interactions among numerous molecules give rise to complexity.

*Developed model for demonstrating the flipping dynamics*

The motion features were also predicted in a developed mathematical model, in which the phase-transition processes with the deformations were included in the above model. The model presented here consists of the combination of the photoreaction kinetic equations (eq (1–3) in the Methods section), the Landau mean-field formulation for the first-order phase transition, and elastic interactions inside the object, and it represents the dynamics well, as shown below. The hierarchical structure of the model is explained in Fig. 8. A two-dimensional crystal was regarded as an assembly of crystalline domains consisting of primary units of the crystalline phase, the phase transition of which was triggered by the photoisomerization of azobenzene and by the cooperative effects derived from the state of the surrounding crystalline phase. The *cis*-isomeric ratio for trigging the first-order phase transition of the unit was defined as $\chi_{PT}$, and the phase transitions occurred after $\chi$ of the unit crosses the value $\chi_{PT}$. The cooperative factor was considered according to the Landau mean-field formulation: the ratio of the meta-phase units in a domain ($\phi_i$) can be expressed as eq (14) with a diffusion term. Fig. 9a shows an example of the functions for the phase stability under a self-oscillatory condition, and the trajectory of ($\chi_i$, $\phi_i$) is drawn in Fig. 9b. According to the model, we hypothesized that each domain has a spontaneous curvature proportional to the $\phi$ value, and we also employed elastic interactions between domains to demonstrate the mechanical deformation behavior (eq (15)).

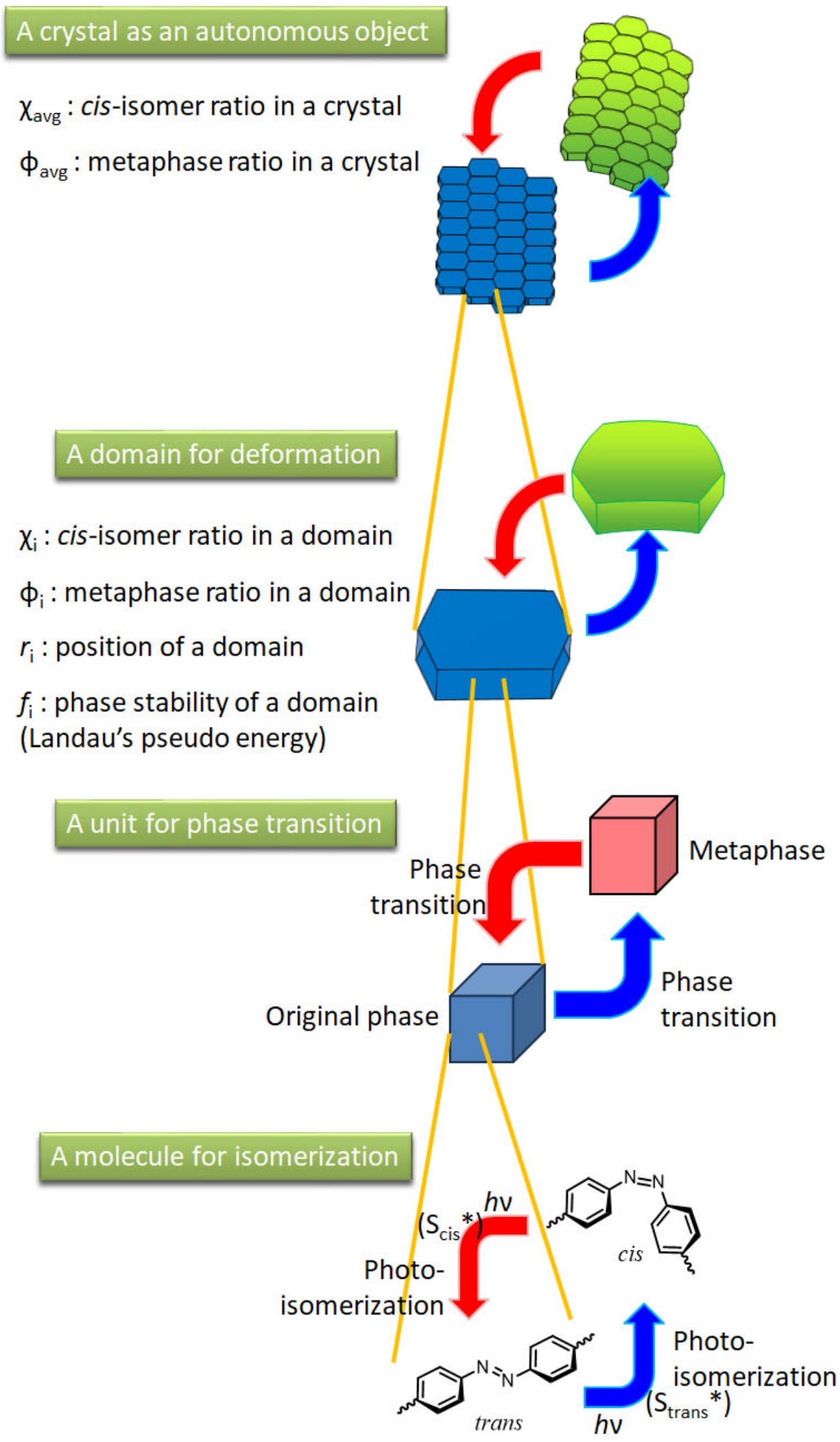

**Figure 8.** Hierarchical structure of the molecular system assumed in the mathematical model. Each domain (*i*) is treated as a subsystem, and the variation in the ratio of the original phase and the metaphases within it ($\phi_i$) is determined using the mean-field model inspired by Landau's first-order phase-transition model, with $\phi$ as the order parameter and the *cis*-isomeric ratio ($\chi_i$) as the trigger parameter. Additionally, the assumption that the spontaneous curvature of the domain is proportional to $\phi$ is introduced, and the deformation of the higher-order structure, the crystal, is modeled through elastic interactions between domains.

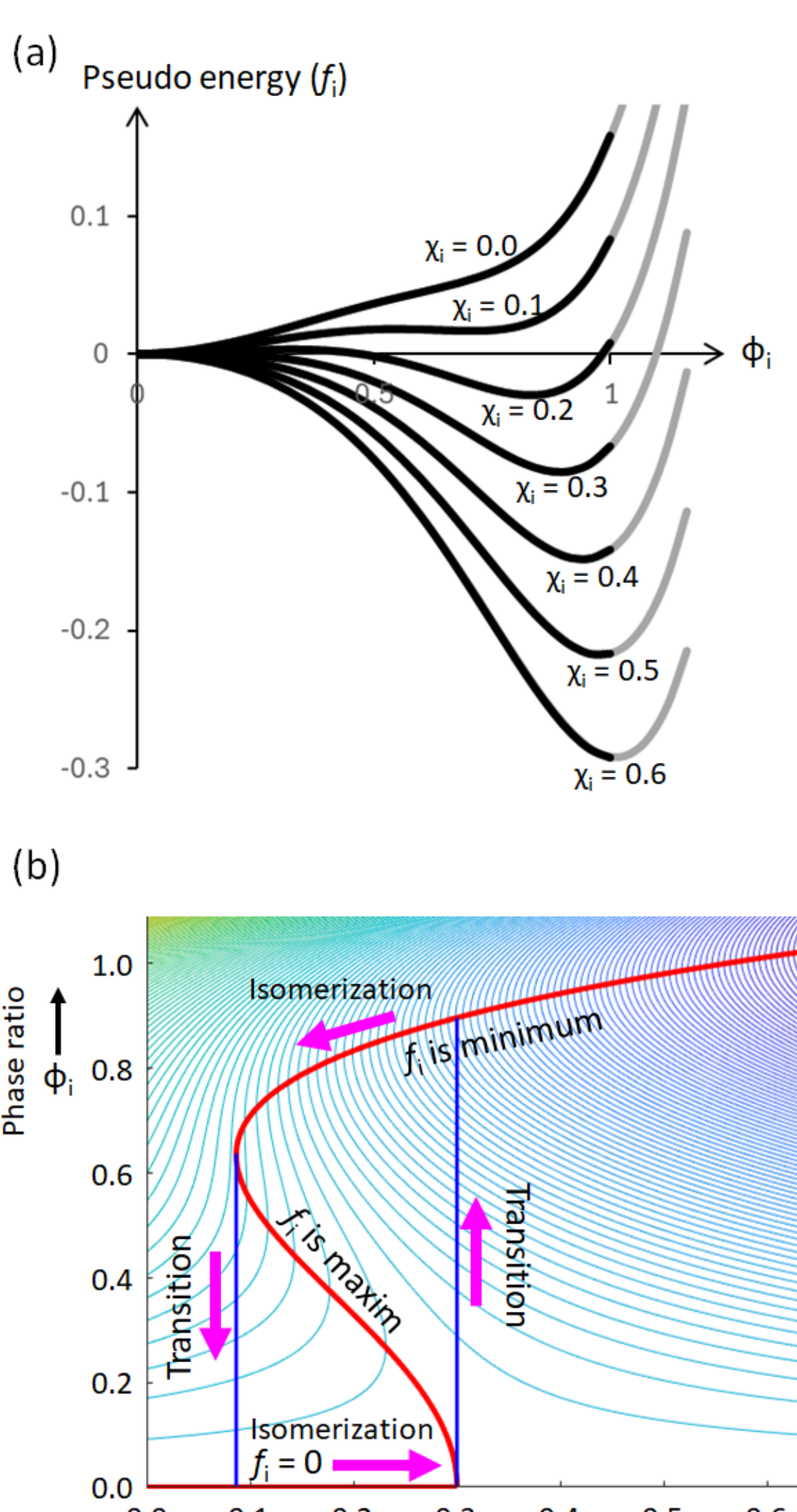

**Figure 9.** Shape of the pseudo energy function shown as eq (9), where $\phi_i$ is defined in the range of 0 to 1. (a) The cross section of the contour graph of $f_i$ (b). While $\chi_i$ is between 0 and 0.086, the minimum of the function is at $\phi_i = 0$. On the other hand, when $\chi_i$ is greater than 0.3, the function has a minimum around $\phi_i = 0.9$. When $\chi_i$ lies between these values, the function is of the double-minimum type, and the phase state depends on the reaction history.

Fig. 10 and Movie S6 show examples of the behaviors calculated by the model. The parameter differences among the models shown in Fig. 10 are only in the outline of the crystal, while all other parameters, including the kinetic coefficients, are the same. We successfully reproduced the oscillatory flipping motion of each object visually with indication of the changes in the *cis*-isomeric ratio and the metaphase fraction. Furthermore, the model indicated that the oscillation frequency depends on the physical shape of the crystal. In the mathematical model, the primary factors causing differences in oscillation periods due to the crystal shape included variations in the degree of deformation resulting from differences in the effect of elastic interactions and differences in changes in the light-receiving area associated with deformation depending on the shape. In addition, the slower deformation could not follow the change in the chemical composition. In the actual system, the shape exhibited even greater diversity, and cooperative mechanisms such as synchronization between oscillators also played a role. In other words, a mathematical model like this should not be expected to fully predict the spatiotemporal patterns of each crystal. However, the analysis of the mathematical model provides us with insights into the characteristics of the phenomena.

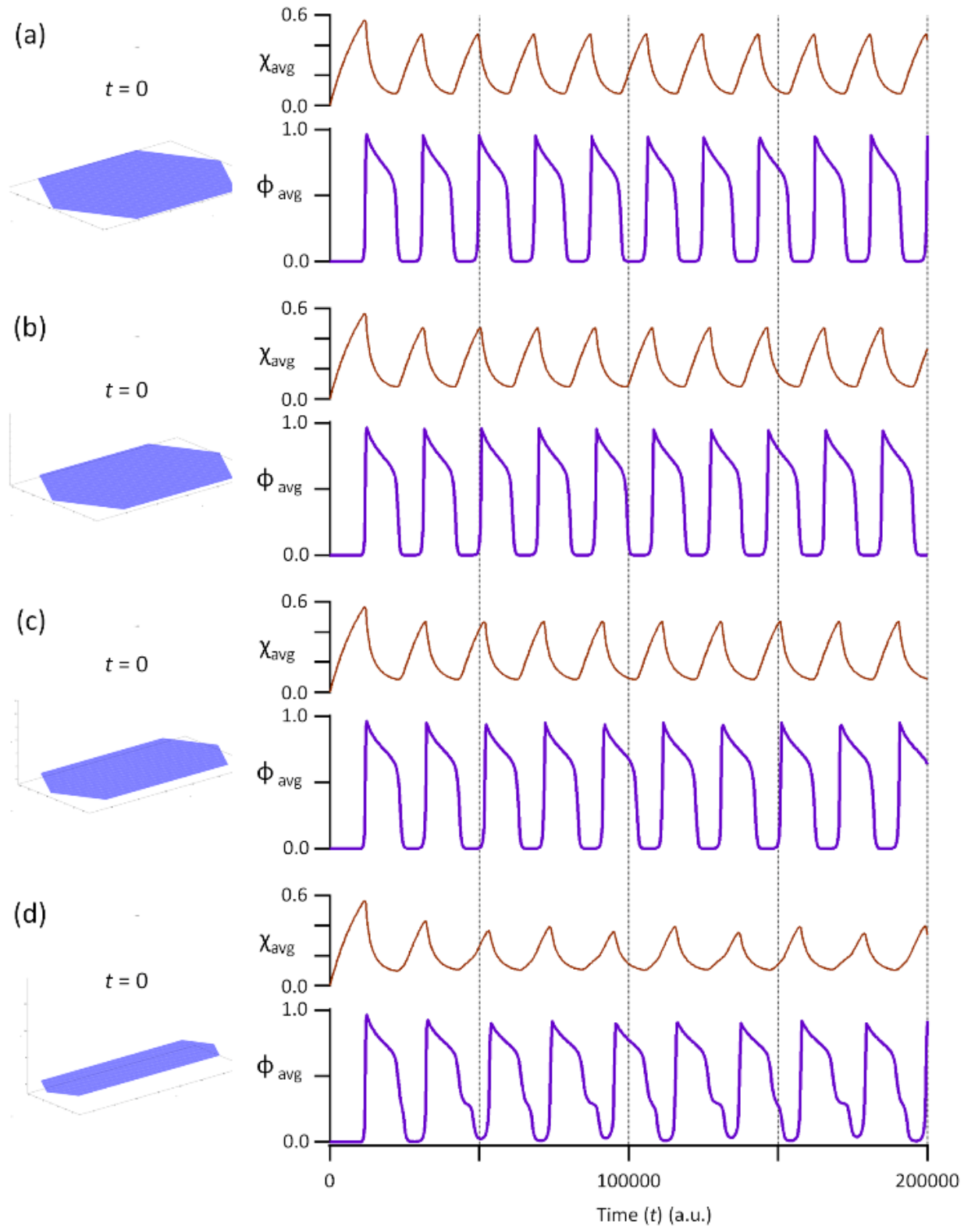

**Figure 10.** Differences in the time profiles in the calculated results due to variations in the outline shape. The calculations were performed with progressively elongated shapes from panels (a) to (d), while keeping all other parameters identical. As the shape became more elongated, the oscillation period increased. As shown in panel (d), the motion continued without returning to a planar shape and transitioned into a stable periodic motion after a computation time of $3 \times 10^5$ a.u. (Movie S6 in ESI).

One of the successes of the model is that behaviors that are different from those observed in stimulus-responsive materials were well visualized in the model (Fig. 11). In planar crystals, although the system initially started with a higher *trans*-isomeric ratio, the state with a higher *cis*-isomeric ratio persisted for a longer duration. Moreover, at the time of the start of deformation to the curved shape followed by the phase transition, the *cis*-isomeric ratio started to decrease, and the state with a small *cis*-isomeric ratio continued in the curved crystal. Then, immediately after the phase transition to the original phase, the *cis*-isomeric ratio started to increase and deformation of the crystal to the planar shape occurred (Movie S7 in ESI). This behavior, with a time lag between the molecular-level change and the macroscopic-level change, intrinsically differed from that of stimulus-responsive materials, of which the molecular-level state could be explained to be almost linked to the deformation state of the material. [54]

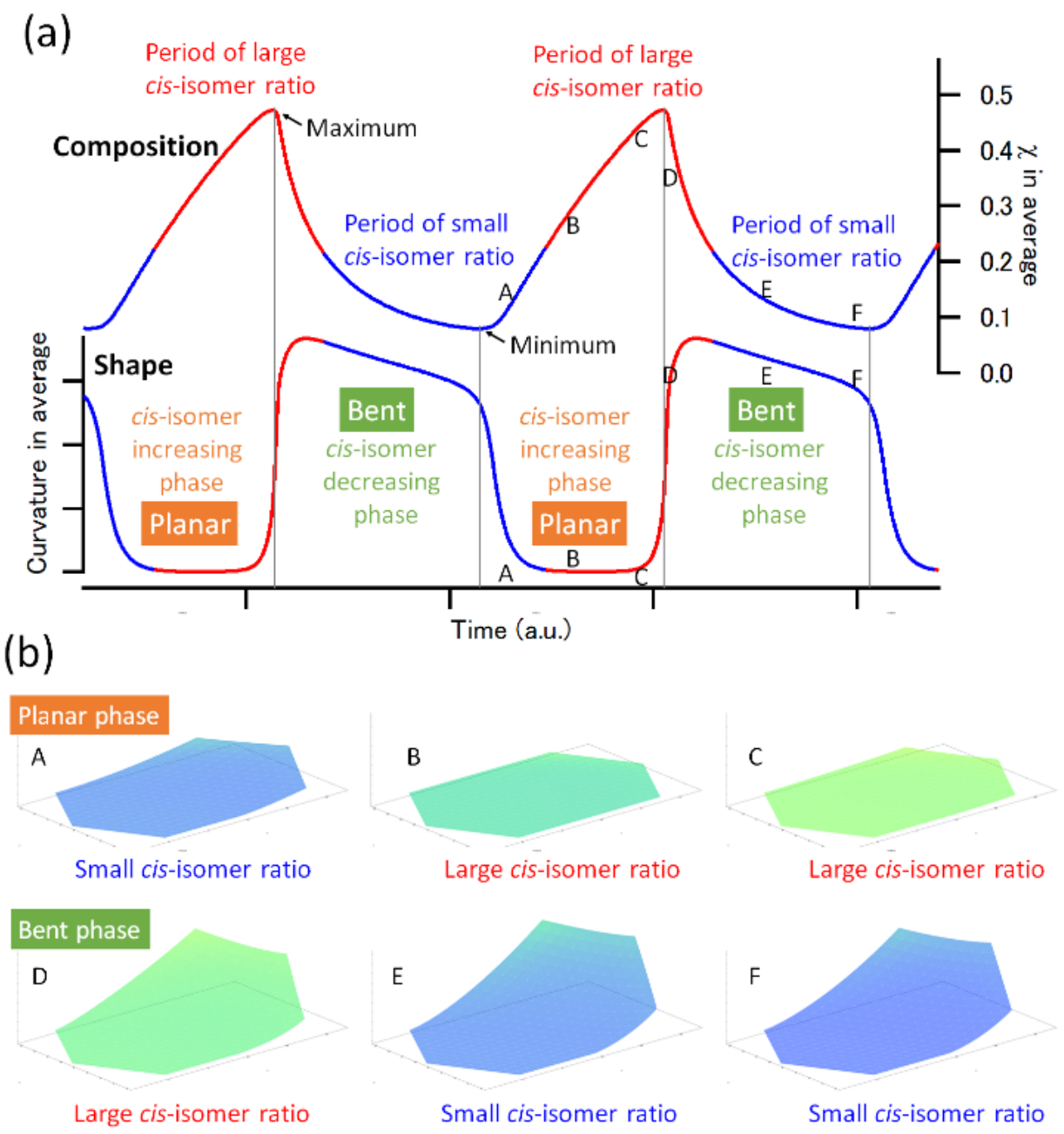

**Figure 11.** Analysis of the mathematical model results shown in Fig. 10b. (a) The time profile of the composition and curvature of the crystal. Red and blue indicate that the *cis*-isomeric ratio ($\chi_{avg}$) is greater or less than 0.22, respectively. (b) Snapshots of the calculated shapes of the crystal at different *cis*-isomeric ratios according to the time profile (A–F). The colors represent the χ-value (blue, green, and greenish yellow indicate ∼0.1, ∼0.3, and ∼0.5, respectively). The results indicate that the shape does not directly depend on the chemical composition; rather, the shape influences the variation in composition.

*Another hysteresis due to the history of the polarized-light rotation*

Next, we experimentally measured the flipping frequency of the crystal used for the measurements in Fig. 4 while rotating the polarization angle by 5° every 5 s, without exposing the crystal to unpolarized light during the experiment. As shown in Fig. 12a and Fig. S5 in ESI, the oscillation frequency differed depending on whether the polarization angle was rotated clockwise or counterclockwise. Similar behavior was also observed in another crystal, the polarized light dependency was four-leaf type (Fig. 13). The behavior is unexplainable from the models shown above. The presence of a molecule or a subsystem that stores the previously received information is required.

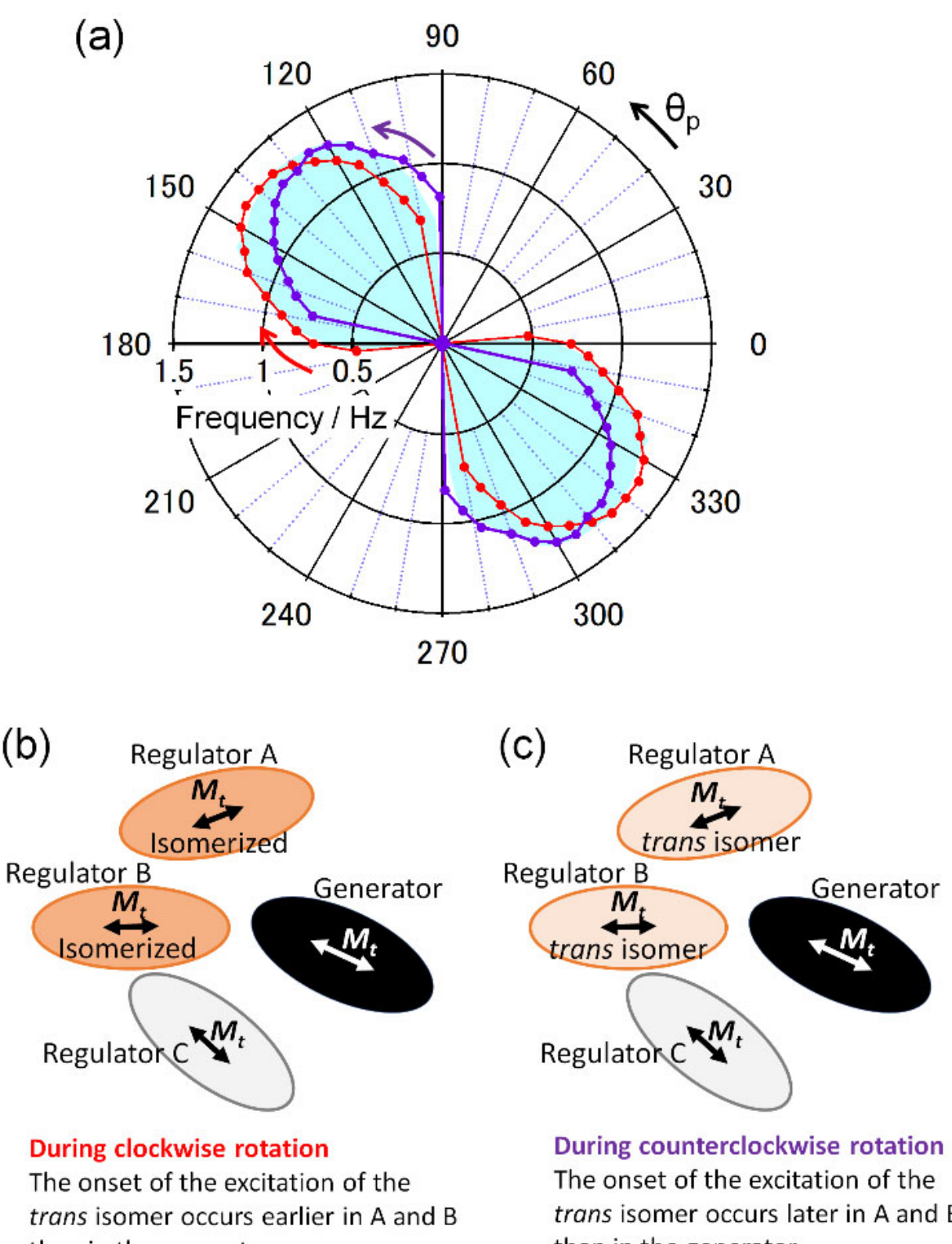

**Figure 12.** (a) Frequency of the self-oscillation of the crystal shown in Fig. 4 when the polarization angle of the blue light was changed stepwise in increments or decrements of 5° (for details, see Movies S3 and S4 in ESI). The red line and the purple line represent the results for clockwise and counterclockwise rotations, respectively. The cyan-colored area indicates the results shown in Fig. 4. (b and c) Schematic illustrations explaining the hysteresis observed with the polarization rotation of the excited light. In the crystal, azobenzene molecules (represented as ellipses) of different symmetries are present, each with a transition moment in a different direction (represented as arrows in the ellipses). The rotation direction of the polarization angle determines the order in which the azobenzene molecules are photoexcited, causing the state of the crystal to vary with the rotational direction. In short, when the generator's transition dipole moment aligns with the polarization direction, the crystal components differ between (b) and (c) due to the difference in the rotational direction of the polarized light. Additional data are shown in Fig. S5 in ESI.

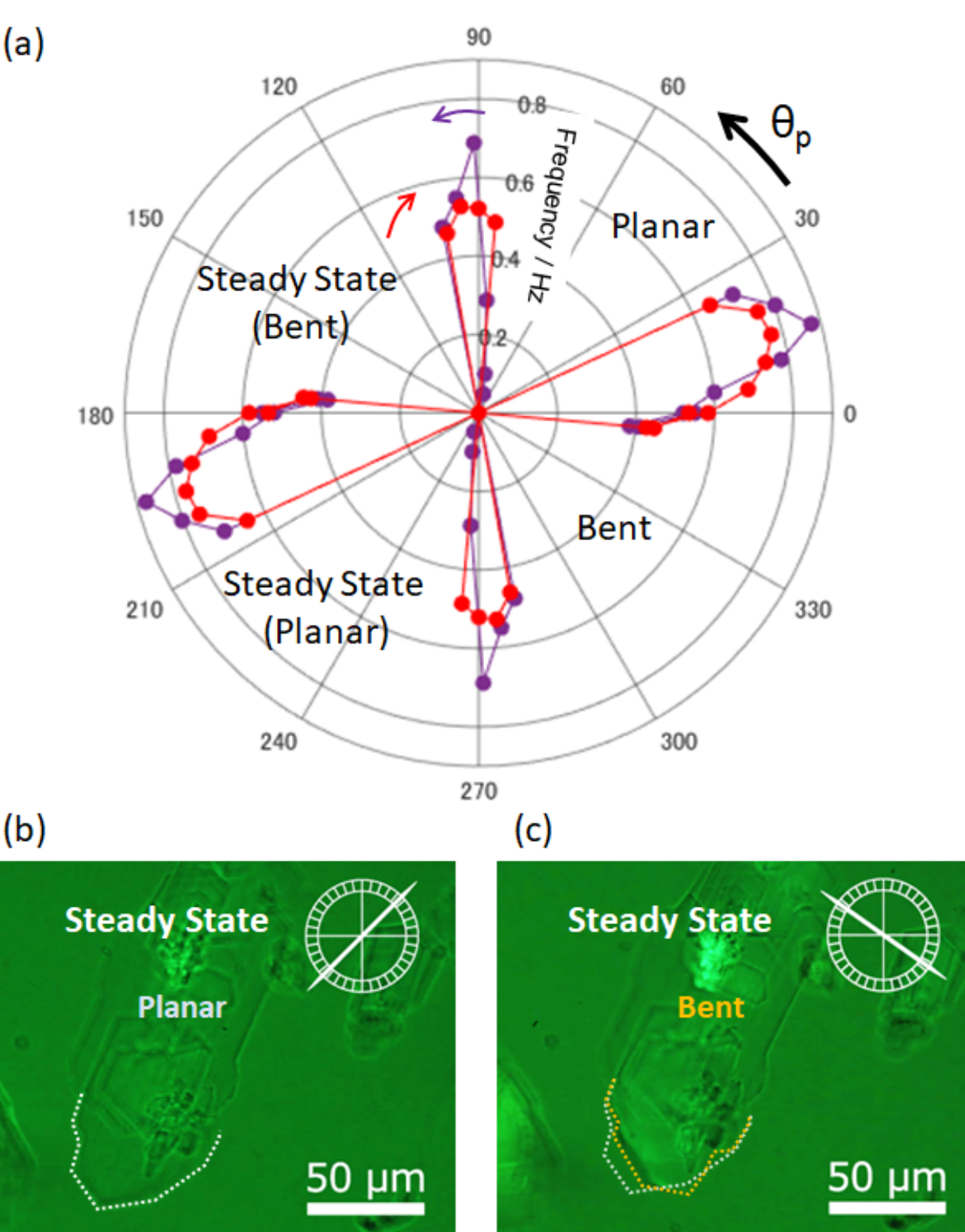

**Figure 13.** (a) Frequency of the self-oscillation of a crystal when the polarization angle of 435-nm light was changed stepwise (Movies S9 and S10 in ESI). The red and purple lines represent the results for clockwise and counterclockwise rotation, respectively. (b, c) Micrographs of the corresponding crystal under 435-nm-polarized-light irradiation, the angles of which were 45° and 135°, respectively. Under these conditions, the crystal did not show self-oscillation but formed a planar shape and a bent shape, respectively. Under oscillation conditions, the crystal deformed between the two shapes. The dashed lines are for easy visualization: white outlines the planar shape, and yellow outlines the bent shape. The micrographs appear green due to the use of a green filter to block the blue light in the observation light.

Based on the crystal structure reported in our previous paper (Fig. 1),[47] we hypothesized that the *cis*-isomeric ratio of the regulator molecules, which varies with the polarized light supplied, changes the frequency of the self-oscillation driven by the generator molecules. The direction of each transition moment of the regulator azobenzene molecules and the generator azobenzene molecules is different. Therefore, the rotational direction of the polarization angle determines whether the regulator molecules begin to accept the light polarization efficiently before or after the generator molecules do (Fig. 12b and c). This is caused by a difference in the *cis*-isomeric ratio of the regulator molecules, which depends on the history of the rotational direction and consequently changes the characteristics of the crystal. If we regard the *trans* isomer and the *cis* isomer as different molecules, the crystal obtained after clockwise versus counterclockwise rotation of polarized light is one with different components, despite the fact that the object is identical. Thus, the frequency necessarily varies with the rotational direction of the polarization.

However, to directly confirm these hypotheses, it would be desirable to quantify the *cis*-isomeric ratio of molecules with specific orientations in the crystal. Nevertheless, given the currently available analytical techniques, it is difficult to carry out such measurements. Despite these limitations, experimental observations support the hypothesis. As evidenced by the results and as expected based on the above discussion, the frequency of self-oscillation generally reflects the history of an external operation, i.e., the rotation of the polarization angle of the excitation light. Indeed, hysteresis was observed in all cases, despite the variations in its expression from crystal to crystal (Fig. S2 and S6 in ESI). Furthermore, as shown in Fig. S2e-h, the presence of hysteresis was reproducible, even though the specific behavioral phenotype was not always identical. For completeness, we note that the idea that the presence of molecules with different crystallographic symmetries contributes to the emergence of hysteresis in response to the polarization rotation is not limited to the discussion of a single-crystal phase. For example, the existence of twin structures also implies the presence of molecules with orientations different from those of the generator molecules in the targeted phase, which could similarly underlie the hysteresis and the resulting behavioral complexity.

*Memory effects on the patterned dynamics*

The memory effect due to the polarization-light hysteresis also modulates the self-oscillatory behavior under unpolarized light irradiation. Fig. 14 (Movie S11 in ESI) and Fig. 15 (Movie S12 in ESI) show the self-oscillation behavior under unpolarized light irradiation following a 2-s irradiation of polarized light. As shown in Fig. 14, the frequency became slow during the first 4 s only when the preillumination of 165°-polarized light was applied, and it recovered by keeping the light irradiation. This finding indicates that the behavior was influenced by the prestimulus, albeit for a short period of time. For a crystal with a complex structure, we also found that the temporal pattern was modulated by the preillumination of 470-nm polarized light. An example is shown in Fig. 15. If the angle of the preilluminated polarized light was 90°, 150°, or 180°, the crystal repeatedly changed into five shapes, [A→B→C→D→E→], after induction periods under unpolarized 435-nm light. On the other hand, if the angle of the preilluminated polarized light was 120°, the crystal repeated the change among three shapes, [F→G→H→], after an induction period under unpolarized 435-nm light. Here, we do not discuss whether forms A, C, and F; forms B, D, and G; and forms E and H are the same structures or not, although the outlines are similar.

Herein, we present two different examples of modulation by preillumination with polarized light. In both cases, as well as in other experiments, the preillumination polarized light modulation on subsequent dynamics was observed when the polarization angles were between 20° and 25° relative to the long side of the crystal. This fact supports the aforementioned assumption that preisomerization of a specific azobenzene molecule in the crystal plays a key role in the modulation.

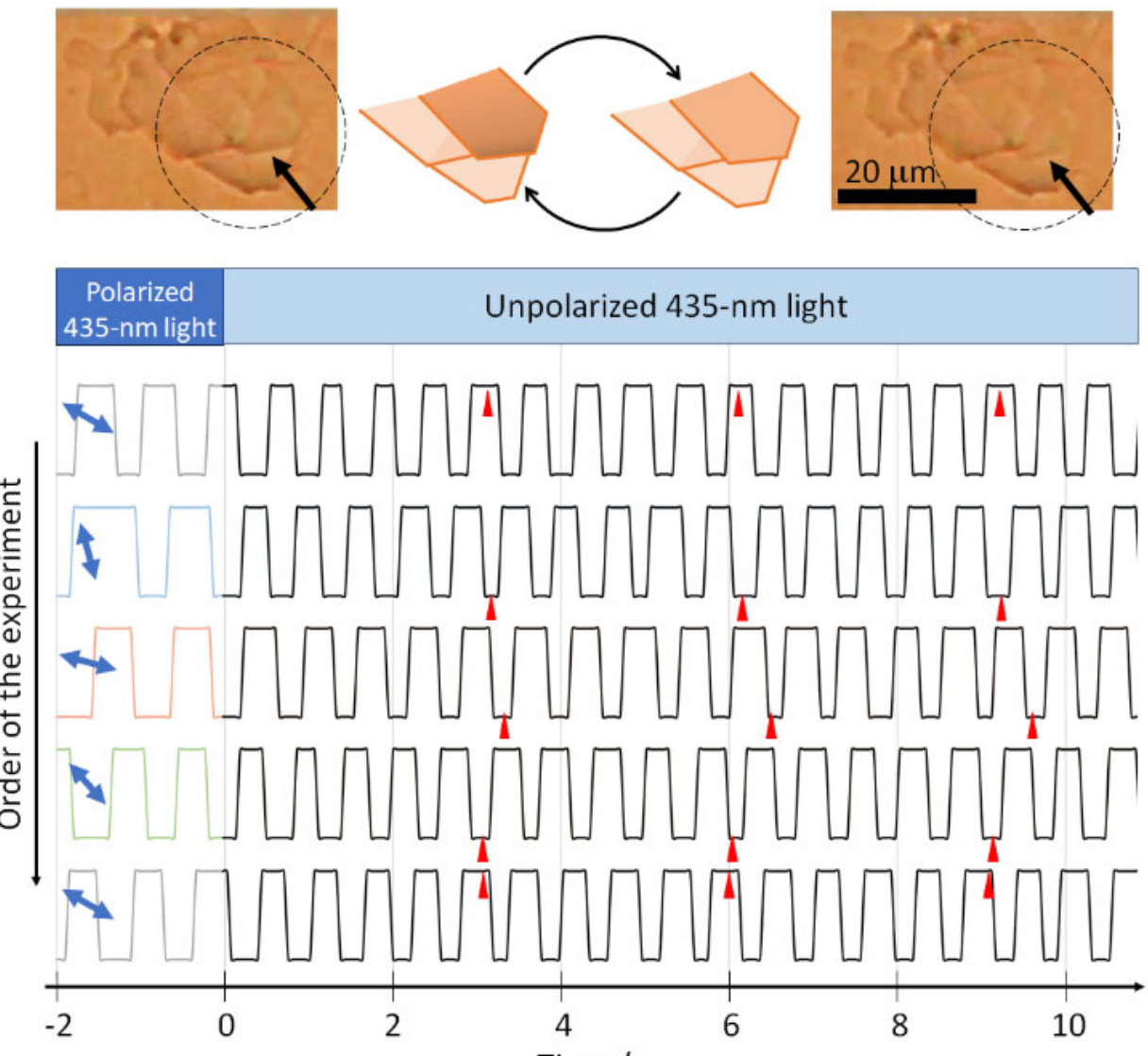

**Figure 14.** Self-oscillation patterns of a crystal under unpolarized light irradiation after being exposed to polarized light for 2 s. The patterns in the graph represent the binarized brightness of an area of the crystal, and the raw data are shown in Movie S11 and Fig. S7 in ESI. Blue arrows indicate the angle of the polarized light (from top to bottom: 150°, 105°, 165°, 135°, and 150°, respectively). Red triangles are shown every five cycles of each oscillatory pattern. The direction of the long side of the flipping crystal in the picture is from the top left to the bottom right.

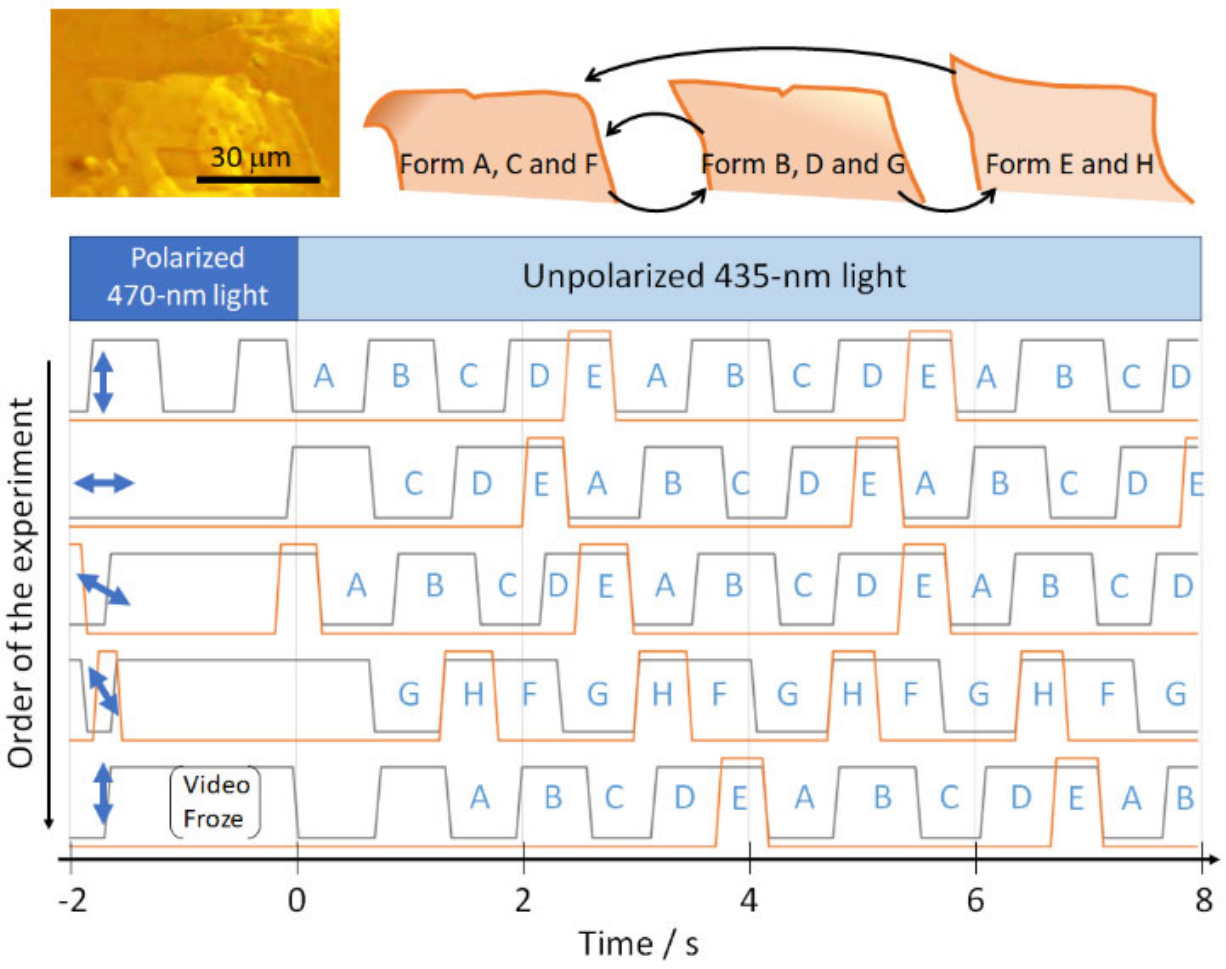

**Figure 15.** Self-oscillation patterns of a crystal under unpolarized light irradiation after being exposed to polarized light for 2 s. The patterns in the graph represent the binarized brightness or brightness change in two areas of the crystal, and the raw data are shown in Movie S12 and Fig. S8 in ESI. Blue arrows indicate the angle of the polarized light (from top to bottom: 90°, 180°, 150°, 120°, and 90°, respectively). The direction of the long side of the flipping crystal in the picture is from the top left to the bottom right.

## Conclusions

In this paper, we analyzed the self-oscillation of a crystal under blue-polarized light irradiation, verified the validity of the inferred reaction mechanism, and constructed a four-dimensional mathematical model based on the proposed mechanism to study the hierarchical self-organized behavior characteristics exhibited by light-driven self-oscillating crystals. The results revealed that the frequency of light-driven self-oscillation varies with both the polarization angle of the incident light and its rotational direction. Since the self-oscillation arose from multiple photoexcitation processes, the range of polarization angles over which self-oscillation occurs differed from the typical conditions for photochemical processes: the dependence on the polarization angle exhibited either a two-leaf or four-leaf pattern in the experimental polar plots, and this behavior was successfully explained theoretically. These distinct patterns correspond to whether the closed trajectory in the phase diagram intersects the self-oscillation region once or twice. Importantly, this theoretical framework suggests that light-driven self-oscillation is a more generalizable phenomenon, potentially occurring in a wide range of molecular-oriented assemblies that undergo photoisomerization-induced phase transitions ‒a well-established phenomenon in chemistry.[55-59]. Whereas the state on the phase diagram is a specific point depending on the light wavelength under unpolarized light irradiation, the state shifts from left-end to right-end and from bottom-end to top-end on the phase diagram depending on the polarized light angle when the light is polarized. By selecting appropriate wavelength and polarized light angle, self-oscillatory behavior can be realized.

Conversely, our crystal demonstrated stable self-oscillation under unpolarized blue light. This stability was disrupted by the light polarity; and under certain conditions, oscillation ceased. This finding contrasts with many other stimulus-responsive materials, where anisotropic deformation is induced when energy is supplied with directional bias.

Although the experimentally observed dynamics appear to be complex, we still believe that the fundamental mechanism underlying self-oscillation remains simple, as depicted in Fig. 1. Indeed, a model derived from this proposed mechanism was able to closely reproduce the observed self-oscillatory behavior. On the other hand, due to the interplay between chemical processes and physical properties, a complete prediction of the spatiotemporal patterns is inherently difficult, and determining precise parameter values remains an unresolved challenge. Even if the underlying chemical processes at the molecular level were fully elucidated, it would still be impossible to completely predict the system's behavior: small fluctuations and differences in kinetic parameters, molecular alignments, or threshold values cause significant differences in behavior. This limitation is likely an intrinsic characteristic of autonomous molecular systems, which operate in close association with chaotic systems.

Throughout this theoretical investigation, we assumed that the chemical reactions between the two isomers proceed in both the forward and reverse directions, although the system does not reach equilibrium. While the photoisomerization reaction proceeds through an irreversible process via the excited state, we did not attribute the emergence of light-driven self-oscillation to this irreversibility. Instead, the system is kept under a steady energy flow, within which a phase transition occurs before the intrinsic chemical reactions reach equilibrium. This phase transition alters the reaction rates, shifting the equilibrium point. By repeating this process, a chemical reaction system that never reaches equilibrium is realized. Even though photochemical reactions were employed in the experiments, the phenomena observed here could also occur in thermally driven chemical reactions, and the constructed model can be applied to such thermal systems as well.

Furthermore, since the information on light polarization is stored as the isomeric composition of the regulatory molecules, the spatial and temporal patterns of self-oscillation were found to depend on the sequence of the external light-irradiation operations. These results indicate that the memory of external operations is retained at the molecular level, modulating the autonomous dynamics of the crystal. In this process, molecules other than those actively driving oscillation—namely, nonoscillatory molecules within the crystal or those in the twin structure, or both—are inferred to play a role in memory and modulation. This concept contributes to the realization of short-term memory in life-like autonomous materials. These findings not only highlight a molecular-level mechanism for short-term memory in autonomous systems but also suggest that autonomous systems and biological systems share fundamental design principles. Just as biological organisms rely on the cooperation of active molecular motors and passive sensors, our study demonstrates how self-governing behavior and adaptive modulation can emerge from molecular systems under nonequilibrium conditions.

## ASSOCIATED CONTENT

The supporting information is available free of charge via the Internet at http://pubs.acs.org.
Figures S1–S8 (PDF)
Movies S1–S17 (MP4)

## AUTHOR INFORMATION

### Corresponding Author

**Yoshiyuki Kageyama** – Faculty of Science, Hokkaido University, Kita-10 Nishi-8, Kita-ku, Sapporo, Hokkaido, 0600810, Japan. ORCID: 0000-0002-2287-832X; Email: y.kageyama@sci.hokudai.ac.jp

### Authors

**Yasuaki Kobayashi** – Department of Mathematics, Faculty of Science, Josai University, Keyakidai, Sakado, Saitama, 350-0295, Japan. ORCID: 0000-0003-4966-963X; Email: ykobayashi@josai.ac.jp
**Makiko Matsuura** – Faculty of Science, Hokkaido University, Kita-10 Nishi-8, Kita-ku, Sapporo, Hokkaido, 0600810, Japan.
**Toshiaki Shimizu** – Faculty of Science, Hokkaido University, Kita-10 Nishi-8, Kita-ku, Sapporo, Hokkaido, 0600810, Japan.

**Norio Tanada** – Faculty of Science, Hokkaido University, Kita-10 Nishi-8, Kita-ku, Sapporo, Hokkaido, 0600810, Japan.
**Daisuke Yazaki** – Faculty of Science, Hokkaido University, Kita-10 Nishi-8, Kita-ku, Sapporo, Hokkaido, 0600810, Japan. Current Address: Faculty of Engineering, Hokkaido University, Kita-13 Nishi-8, Kita-ku, Sapporo, Hokkaido, 0600813, Japan.

## Author Contributions

Y. Kageyama designed and performed the experimental studies and mathematical analysis for the reaction kinetics and prepared the manuscript. Y. Kobayashi designed the mathematical four-dimensional model, and T.S. and Y.K. optimized the model and performed the calculations. M.M., N.T., and D.Y. assisted with the experiments, and M.M. also supported the movie analysis.

## Notes

The authors declare no competing financial interest.

## ACKNOWLEDGMENT

This work was supported by JSPS KAKENHI grant JP18H05423 within the Scientific Research on Innovative Area "Molecular Engine," JP 20H04622 within the Scientific Research on Innovative Area "Math Materials" and JP23H04392 within the Transformative Research Area "Molecular Cybernetics" to Y. Kageyama and JP22K03428 under Grant-in-Aid for Scientific Research (C) to Y. Kobayashi. The authors thank Write Science Right for providing English language editing.

TOC-Graphic

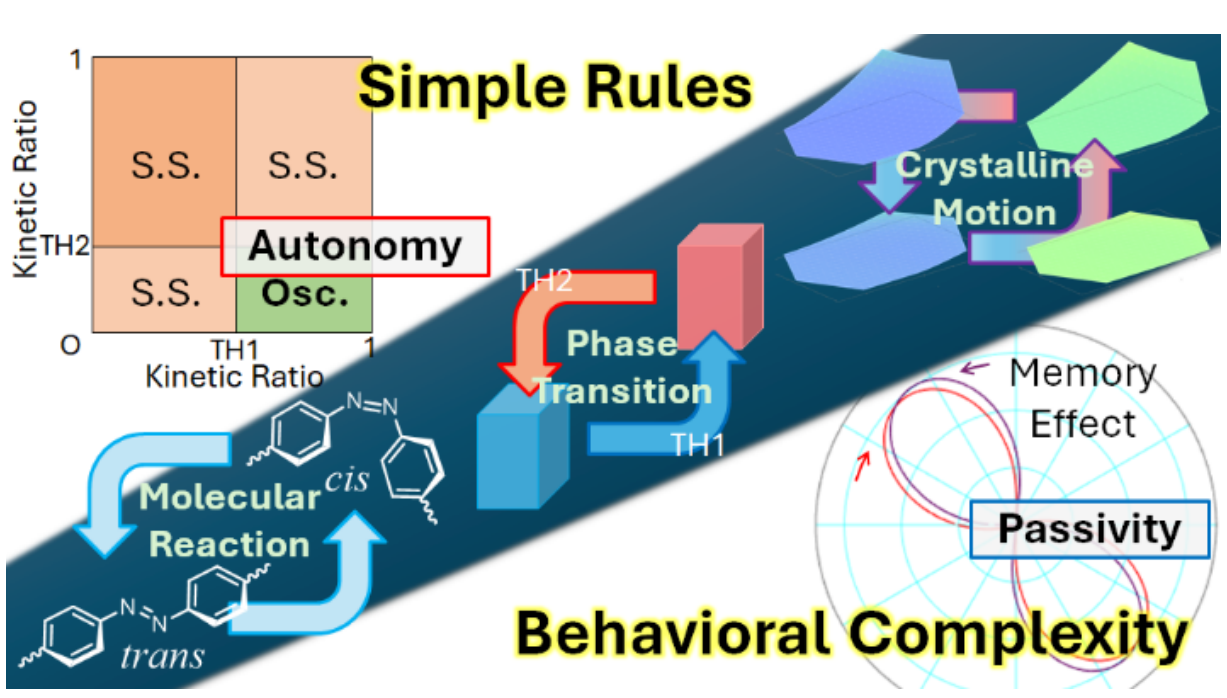